\documentclass{elsart}
\usepackage{epsfig}
\begin{document}
\begin{frontmatter}
\title{Advanced Compton Telescope Designs \\
 and SN Science}

\vspace{-6mm}
                                                
\author[c1]{P.A.~Milne}
\author{R.A. ~Kroeger}
\author[c3]{J.D.~Kurfess}
\author[c3]{L.-S.~The}
\address[c1]{NRC/NRL Resident Research Associate, Naval Research Lab,
Code 7650,Washington DC 20375}
\address[c2]{Naval Research Lab, Code 7650,Washington DC 20375}
\address[c3]{Clemson University, Clemson, SC 29631}

\vspace{-3mm}

\begin{abstract}

\small
The Advanced Compton Telescope (ACT) has been suggested to be the optimal 
next-generation instrument to study nuclear gamma-ray lines. In this work, 
we investigate the potential of three hypothetical designs of the 
ACT to perform SN science. 
We provide estimates of 1) the SN detection rate, 2) the SN Ia discrimination rate, 
and 3) which gamma-ray lines would be detected from specific supernova remnants.
We find that the prompt emission from a SN Ia is such that
 it is unlikely that one would be within the 
range that an INTERMEDIATE ACT would be able to distinguish between explosion 
scenarios,
 although such an instrument would detect a handful of SNRs.
 We further find that the SUPERIOR ACT design would  be a truly breakthrough
instrument for SN science. 
By supplying these estimates, we intend to assist the gamma-ray astrophysics 
community in deciding the course of the next decade of gamma-ray SN science. 
\end{abstract}

\end{frontmatter}

\vspace{-12mm}
\section{Introduction}

\vspace{-6mm}
Detections of gamma-ray lines from supernovae are some of 
the most exciting aspects of ``Astronomy with Radioactivites". 
Radionuclei produced in both thermonuclear and core-collapse 
supernovae decay on various timescales leading to potentially 
detectable emission over a range of epochs. Initially, short-lived 
radionuclei produce intense gamma-ray lines that might be observed 
from supernovae in other galaxies. Detection of emission during 
the early epoch permits studies of the evolution of the opacity 
to the gamma-ray emission, and of the yields  of short-lived 
isotopes\cite{clay69}. At later times, longer-lived 
radionuclei produce a less-intense emission that could be 
observed from multiple galactic supernova remnants 
(hereafter call {\em SNR} emission). Detection of 
emission during this intermediate epoch permits the study of the 
yields of these isotopes as well as of the kinematics of the 
SN ejecta\cite{dieh98}. At even later times, long-lived radionuclei produce a 
faint emission that is not attributable to specific remnants, but rather 
is collectively observed as a diffuse emission. Observations of 
this diffuse emission studies the million-year history of supernovae 
in the Galaxy, as a contributor to the total galactic line 
emission\cite{pran96}.
Although interesting, we do not discuss this {\em diffuse} emission.

We simulate the transport of gamma-ray line photons through 
the ejecta of SN models. Combining these simulations with three sets of 
specifications of Advanced Compton Telescopes (ACT), and inferred SN 
rates, we produce estimates of the rate at which supernovae 
will be detectable as {\em prompt} gamma-ray line sources. We also provide  
estimates of the various gamma-ray lines that might be 
detectable from specific supernova remnants. The purpose of this 
exercise is to quantify the expectations of an ACT in performing
 SN science, and thus 
to assist the gamma-ray community in assessing what direction they 
would like to pursue for the future of gamma-ray astronomy.

\vspace{-6mm}
\section{SN Types, Rates and Emission Profile}

\vspace{-5mm}
For this study we assume that {\em 100 SNe Ia per year within 100 
Mpc}. Of these
100 SNe, 60 would be normally-luminous, 20 sub-luminous, and 20
super-luminous.
(The
reader is referred to an earlier work (Milne et al. 2000a) for a derivation
of these rates, and the relevant references\cite{capp97,hamu99,li99}.)  
We further assume that SNe Ia occur 
uniformly through-out space at this density, ignoring the possibility
of a nearby local void. 
The type II/Ib/Ic SN rate has always been suggested to be higher than the 
SN Ia rate (we adopt a rate for the Galaxy of 1.7 SNe II/Ib per 
century\cite{capp00}).  
This is important for studies of {\em SNR} and {\em diffuse} 
gamma-ray line emission. However, the distance to which a type II/Ib 
SN will be detectable is much less than the distance to which a SN Ia 
will be detectable. The much smaller volume that results negates the 
enhanced SN rate. Thus, it is unlikely that a type II/Ib/Ic SN will be 
detected as a source of prompt emission by an ACT.

Considerable variations exist within the collection of SN Ia models.
These variations are driven both by the uncertainty as to the correct
explosion scenario(s) and by the heterogeneity displayed in the spectra
and light curves of observed SNe Ia (for a general discussion, see 
Wheeler 1995).  It is beyond the scope of this 
paper to define and compare SN explosion scenarios, we will simply 
introduce the model types used in this study. 
We have included four single-degenerate Chandrasekhar mass (SDCM) models in
our simulations (all with M$_{ej}$ $\sim$ 1.4 M$_{\odot}$); 
(1) the normally-luminous deflagration, 
W7 (0.58 M$_{\odot}$($^{56}$Ni), Nomoto, Thielemann, \& Yokoi 1984),
(2) the normally-luminous delayed detonation, 
DD23C (0.60 M$_{\odot}$($^{56}$Ni), H\H{o}flich et al. 1998),
(3) the sub-luminous pulsed-delayed detonation, PDD54 
(0.17 M$_{\odot}$($^{56}$Ni), H\H{o}flich, Khokhlov
\& Wheeler 1995), and (4) the super-luminous delayed detonation,
W7DT (0.76 M$_{\odot}$($^{56}$Ni), Yamaoka et al. 1992). 
We have included three variants of the sub-Chandrasekhar mass (SC) 
scenario; (1)
the normally-luminous SC model, HED8 (0.96 M$_{\odot}$ (M$_{ej}$), 0.51
M$_{\odot}$($^{56}$Ni)), (2) the sub-luminous SC model, HED6
(0.77 M$_{\odot}$ (M$_{ej}$), 0.26 M$_{\odot}$($^{56}$Ni),
 H\H{o}flich \& Khokhlov 1996), (3) the super-luminous SC model,
HECD (1.07 M$_{\odot}$ (M$_{ej}$), 0.72 M$_{\odot}$($^{56}$Ni)
Kumagai \& Nomoto 1997).
A double degenerate scenario involving the merger of two CO white dwarfs
has also been suggested to explain SNe Ia (MERG).
We have included three variants of the MERG
scenario; DET2, DET2ENV2, and DET2ENV6 (all with 0.62 M$_{\odot}$($^{56}$Ni), 
and M$_{ej}$ = 1.2, 1.4, 1.8 M$_{\odot}$).
Although the observations of SNe II/Ib also suggest considerable variations 
within this class, for this study we test only the model, 
W10HMM\cite{pint88}. 
Simulations of gamma-ray and positron transport have been performed 
for all of these models. Explanations of the simulation algorithms are  
contained in previous works \cite{burr90,the90,miln99,miln00a}.

After the SN envelope is transparent to gamma-rays,  
line fluxes from SNRs 
 reflect the decay rates, which gives the isotopic yield. In addition, 
the flux levels for Ti44 (1157 keV), Al26 (1809 keV), and Fe60 (1173 keV) 
remain roughly constant for many years, making young and old SNRs potentially 
detectable. However, the intensity of the flux is much lower than for 
{\em prompt} emission, so the majority of detectable SNRs will be 
in the Galaxy.  The SN Ia yields used in this work 
 are primarily from deflagration models (Iwamoto et al. 
1999), and do not treat the variations within the type Ia class. 
The SN II/Ib yields are a compilation as summarized by Knodlseder (2000).
 The r-process 
isotopic yields ($^{126}$Sn) are based upon the assumption that the solar 
abundance of these isotopes has been entirely due to SNe II/Ib occuring at the 
rate of 1.7 SNe II/Ib per century in the Galaxy 
(see Ott, these proceedings).

In addition to these gamma-ray lines, we simulate the 511 keV line emission 
that would be expected from SNRs. For SNe Ia, we assume that 5\% of the 
positrons created in 19\% of $^{56}$Co decays escape the ejecta 
(8 x 10$^{52}$ positrons per SN Ia)\cite{chan93,miln99}. For 
SNe II/Ib, we assume that 100\% of $^{44}$Ti and $^{26}$Al decay positrons 
(scaled to 0.95 and 0.82 branching ratios, respectively) escape the 
ejecta. In both cases we make the {\it ad-hoc} assumptions 
that 1) the escaping positrons annihilate on a 10$^{5}$ year 
timescale\cite{gues91}, and 2) the annihilation radiation matches 
that observed from the Galaxy-wide emission (i.e. 
 with 0.58 511 keV photons produced 
 per annihilation (f$_{Ps}$=0.95), and the 511 keV line having a 5 keV 
FWHM\cite{harr98,kinz01}). These assumptions lead to fluxes of 
1.2 x 10$^{-4}$ phot cm$^{-2}$ s$^{-1}$ and 
3.7 x 10$^{-6}$ phot cm$^{-2}$ s$^{-1}$, respectively, 
from 1 kpc distant SN Ia and  SN II/Ib remnants.

{\em Prompt} SN emission has been treated as being emitted by a 
point source for the various instruments, but for {\em SNR} emission
the angular sizes of the larger remnants are important. We 
performed our estimates assuming two extremes.  The 
``extended source" extreme assumes that all of the emission is 
uniformly distributed throughout the projected optical remnant. 
We then scaled the sensitivity such that the effective sensitivity 
scaled as, 
{\it Extended Sensitivity}  = {\it Point Source Sensitivity} 
$\cdot$ ( $\theta_{SNR}$ 
/ $\theta_{Det}$ )$^{1/2}$,  
where $\theta_{SNR}$ is the angular size of the optical remnant, and 
$\theta_{Det}$ is the angular resolution of the detector. This is 
the lower limit for the detection of gamma-ray lines used in this 
work.  The ``single knot" 
extreme assumes that all of the emission emanates from a single knot 
that is smaller than the detector's angular resolution. This is 
clearly an upper limit for the detection of gamma-ray lines.

\vspace{-8mm}
\section{Compton Telescope Options and Mission Scenarios} 

\vspace{-5mm}
NASA's Gamma-Ray Working Group (GRAPWG) identified the study of
nuclear astrophysics and sites of gamma-ray line emission as its
highest priority science topic (June 1999), and the development of an Advanced
Compton Telescope (ACT) as its highest priority major mission.
The GRAPWG has outlined a baseline ACT with a goal of achieving a
point source localization accuracy of
of $\sim$ 5', an energy resolution of $\leq$ 3 keV
(@1 MeV), a FoV of 60$^{\circ}$ and a broad-line sensitivity of
1 x 10$^{-6}$ phot cm$^{-2}$ s$^{-1}$ (10$^{6}$s, 3$\sigma$).
The Naval Research Laboratory (NRL) is investigating both germanium
 and silicon Compton telescope designs,
with the intention of exceeding the baseline
specifications in both sensitivity and FoV (Kurfess et al. 1999).
The advances that will make
this improvement possible are: 1) large volume detector arrays which will
increase the effective area, 2) excellent spatial and energy
resolution from the use of position-sensitive solid-state detectors, and
3) in one mode of analysis, 
 employing two Compton scatters and a third interaction
 to determine the incoming energy and
angle rather than a single scatter and total energy absorption.  
The current NRL/ACT design will
increase the FoV to 120$^{\circ}$, and improve
the broad-line sensitivity to 3 x 10$^{-7}$ phot cm$^{-2}$ s$^{-1}$
(10$^{6}$s, 3$\sigma$). The point source localization
will be poorer than the
baseline, increasing to 30' -60' for the weakest sources, but this 
improves for strong sources. The energy resolution is expected to
be $\sim$ 20 keV. 

It is generally accepted that the timescale to construct an ACT capable of 
meeting either the GRAPWG specifications, or the NRL specifications, 
will be more than ten years. It has been argued that an ``intermediate" 
instrument needs to be constructed, both to continue to develop the 
relevant technologies involved with the construction of an ACT, and to 
provide some data for scientists working in this field. During this 
meeting Bloser desribed one design of an intermediate 
ACT, the Medium Energy Gamma-ray Astronomy ACT. We have performed 
simulations loosely based upon the specifications described for MEGA. 
Shown in Table 1 are the sensitivities, FoVs, energy and angular resolutions 
that we assume for the ``INTERMEDIATE", ``BASELINE" and ``SUPERIOR" ACT 
concept designs. The accumulation efficiencies depend upon 
the observing mode. We have allowed for two observing modes for these 
designs, surveying or Target of Opportunity (ToO). The survey mode assumes 
that the ACT has been placed in an equatorial orbit. The INTERMEDIATE and 
SUPERIOR instruments are nearly full-sky, and would be zenith-pointed in 
survey mode. The BASELINE instrument would alternate between 
$b$=-30$^{\circ}$, and $b$=+30$^{\circ}$ pointings in survey mode. The 
accumulation efficiency for these instruments is then the combination of 
the fraction of the time the SN would be within the FoV and the estimate 
of roughly 10\% dead-time for each instrument. The
INTERMEDIATE instrument was alternatively operated in ToO mode, responding 
to the optical detection of a SN Ia one week pre-peak, or roughly 11 
days after the explosion.\footnote{Current optical SN searches do not regularly 
achieve -7$^{d}$ detections. However, it is possible that searches will achieve 
such early detections by the time an INTERMEDIATE ACT 
would become operational.} The accumulation efficiency in ToO mode 
was adopted to be 
65\%, combining the influences of earth-occultation and dead-time.

The length of observation affects whether the 
SN or SNR will be detectable in a given gamma-ray line. For {\em prompt} 
emission, we assume that the observations continue until 300$^{d}$ after 
the SN explosion. The survey mode allows the observation of a 
SN from the time of explosion {\it without requiring the optical 
detection of the SN!} For {\em SNR} emission, we assume two year mission 
lifetimes for the survey mode. For ToO mode (better described as 
``pointed mode" when observing SNRs), we assume a two month 
observation. 
Also shown in Table 1 are the same specifications for the INTEGRAL 
instruments SPI and IBIS. These instruments would also operate in 
ToO mode for {\em prompt} emission and pointed mode for SNRs. 
The Vela and RXJ0852-4622 SNRs are scheduled to receive the 
equivalent of a one month observation in the first two years of the 
INTEGRAL Core Program.

\begin{table}
\begin{center}
\caption{{\bf Specifications for three hypothetical ACTs.}} 
\vspace{3mm}
\small
\begin{tabular}{|l|cccc|cc|}
\hline
Spec & INT(Surv) & INT(ToO) & BASE.& SUP.& SPI & IBIS \\
\hline
Broad-line & 1(-5) & 1(-5) & 1(-6) & 
3(-7) & 2.5(-5) & 3.5(-4) \\
\hspace{5mm}sensitivity$^{a}$ & & & & & & \\
Narrow-line & 1(-5) & 1(-5) & 2(-7) &
3(-7) & 1.3(-5) & 5.3(-4) \\
\hspace{5mm}sensitivity$^{a}$ & & & & & & \\
511 keV & 3(-5) & 3(-5) & 4(-6) &
5(-7) & 4.2(-5) & 2.5(-4) \\
\hspace{5mm}sensitivity$^{a}$ & & & & & & \\
FoV & 120$^{\circ}$ & --- & 60$^{\circ}$ & 120$^{\circ}$ & --- & --- \\
Sky Coverage & 87\% & --- & 51\% & 87\% & --- & --- \\
Accum. eff. & 30\% & 65\% & 15\% & 30\% & 65\% & 65\% \\
Ang. Res. & 2.4$^{\circ}$ & 2.4$^{\circ}$ & 5'(b) & 1$^{\circ}$ & 2.5$^{\circ}$ & 12' \\
Energy Res. & 30 & 30 & 2 & 30 & 2 & 60 \\
$[$keV @ 1 MeV$]$ & & & & & & \\
\hline
\end{tabular}
\end{center}
\vspace{3mm}
\small
\begin{tabular}{l}
\vspace{-2mm}
$^{a}$ Sensitivities are in units of phot cm$^{-2}$ s$^{-1}$
(3$\sigma$, 10$^{6}$ s). \\
\vspace{-2mm}
$^{b}$ We note that the angular resolution of ACT designs are 
limited by Doppler \\ 
\vspace{-2mm}
broadening, and it is not clear that 5' angular resolution 
could be achieved.
\end{tabular}
\end{table}

\vspace{-8mm}
\section{SN Science Prospects for Different Intruments}

\vspace{-5mm}
Gamma-ray detection of SNe Ia will be the most basic level of SN science
that will be performed with an ACT. Through SN detections we learn about the
relative rates of SN Ia sub-classes, the SN Ia rate as a
function of galaxy morphological class, and the radial
distribution of SNe Ia. All of these studies assume that the gamma-ray
observations are coordinated with optical observations. SN searches 
suffer from the Shaw effect as well as from extinction effects, as 
highly extincted SNe are less likely to be detected.
Gamma-rays are not expected to suffer from these effects and will detect
all SNe Ia in a galaxy. 

\begin{table}
\vspace{-9mm}
\begin{center}
\caption{{\bf Maximum detectable distance and detection rates for 
various SN models, 
and 
rates at which SN Ia models would be distinguished (5$\sigma$).$^{a}$}}
\vspace{1mm}
\footnotesize
\begin{tabular}{|ll|cc|cc|cc|cc|}
\hline
\multicolumn{2}{|l|}{\bf{Detections}} &&&&&&&& \\  
SN &SN& \multicolumn{2}{c|}{INTER(SURV)} & \multicolumn{2}{c|}{INTER(ToO)$^{b}$}
 & \multicolumn{2}{c|}{BASELINE} & \multicolumn{2}{c|}{SUPERIOR} \\
Sub-Class&Model & Dist & Rate & Dist & Rate & Dist & Rate & Dist & Rate \\
\hline
\multicolumn{2}{|l|}{Normal} &&&&&&&& \\
&W7 & 21 & 0.5 & 25 & 1.0 & 55 & 5.2 & 120 & 90 \\
\multicolumn{2}{|l|}{Super-lum} &&&&&&&& \\
&HECD & 24 & 0.3 & 30 & 0.5 & 65 & 2.8 & 140 & 48 \\
\multicolumn{2}{|l|}{Sub-lum} &&&&&&&& \\
&PDD54 & 11 & 0.0 & 13 & 0.0 & 29 & 0.3 & 63 & 4.4 \\
&HED6 & 15 & 0.1 & 18 & 0.1 & 40 & 0.7 & 86 & 11 \\
&DET2E6 & 17 & 0.1 & 21 & 0.2 & 46 & 1.0 & 100 & 18 \\
\multicolumn{2}{|l|}{Type II/Ib (Distances)} &
\multicolumn{2}{l|}{265 kpc} & 
\multicolumn{2}{l|}{320 kpc} &
\multicolumn{2}{l|}{750 kpc} &
\multicolumn{2}{l|}{1.6 Mpc} \\
\hline
\end{tabular}
\begin{tabular}{ccc}
&&\\
\end{tabular}
\vspace{-3mm}
\small
\begin{tabular}{|llccc|lccc|}
\hline
\multicolumn{4}{|l}{\small{\bf{Discrimination Rates}}} &&&&&\\
\multicolumn{5}{|c|}{BASELINE}  & \multicolumn{4}{c|}{SUPERIOR} \\
\multicolumn{2}{|l}{Normal} & & & & & & &\\
&   & HED8 & DD23C & DET2E2 & & HED8 & DD23C & DET2E2 \\
&W7 & 0.4 & 0.1 & 0.1 & W7 & 6 & 1.4(0.5) & 2.0(1.3)\\
&HED8 & --- & 0.5 & 0.6 & HED8 & --- & 9 & 9.7 \\
&DD23C & --- & --- & 0.2(0.0) & DD23C & --- & --- & 3.2(0.5) \\ 
\multicolumn{2}{|l}{Super-lum} & & & & & & & \\
&  & HECD & DET2 & & & HECD & DET2 & \\
&W7DT & 0.1 & 0.3 & & W7DT & 0.9(0.5) & 4.7(1.8) & \\
&HECD & --- & 0.2 & & HECD & --- & 3.6(2.3) & \\
\multicolumn{2}{|l}{Sub-lum} & & &  & & & & \\
&  & HED6 & DET2E6 & & & HED6 & DET2E6 & \\
&PDD54 & 0.2(0.0) & 0.5(0.0) & & PDD54 & 3.9(0.5) & 8.5(0.4) & \\
&HED6 & --- & 0.3 & & HED6 & --- & 5.4(3.2) & \\
\hline
\end{tabular}
\end{center}
\vspace{-5mm}
\footnotesize
\begin{tabular}{l}
\\
\vspace{-3mm}
$^{a}$ Rates are in units of [SNe/yr]. Detection and 
discrimination are at the 
5$\sigma$ level. \\ 
\vspace{-2mm}
Discrimination rates assume the 
distance to the SN is known and the explosion date \\ 
\vspace{-2mm}
is known to $\pm$ 1$^{d}$. Discrimination rates in brackets () 
 show cases where having the \\
\vspace{-2mm}
SN distance as an unknown appreciably lowers the discrimination 
rate. \\
\vspace{-2mm}
$^{b}$ Assumes that the detector slews to the SN 11 days after the 
explosion. \\
\end{tabular}
\end{table}
\small

Many SNe need to be detected for the estimated SN rates and distributions
to be significant. Shown in the upper panel of 
Table 2 are the distances and rates at which
given SN models could be detected (5$\sigma$) via the combined observation
of the  812, 847, 1238 \& 1562 keV lines. We assume all lines to be detected 
at the quoted broad-line sensitivity, and account for relative intensity.  
For each ACT the $^{56}$Ni-rich super-luminous SNe Ia will
be detected to the largest distances, but the larger SN rate of
normally-luminous SNe Ia make them the most frequently sampled sub-class.
Sub-luminous SNe Ia will be infrequently detected due to both the 
under-production of $^{56}$Ni and the low SN rate. The INTERMEDIATE 
ACT would detect 1-2 SNe Ia per year, and would be slightly more 
efficient at detecting SNe if used as a ToO instrument (but only if 
the telescope can respond to 100\% of nearby SNe Ia more than a 
week before the optical peak). Regardless of the mode of operation,  
a rate of a few SNe Ia per year is inadequate to contribute to the 
study of SN rates. The BASELINE design is more efficient at detecting 
SNe Ia, detecting roughly 10 per year. The SUPERIOR ACT would detect 
almost 160 SNe per year, detecting multiple members of each sub-class. 
We assert that a 5 year sampling of 800 SNe Ia will allow SN rate 
studies capable of quantifying the biases of optical searches.

For a subset of the detected SNe Ia, the line fluxes will be large
enough to generate gamma-ray light curves. 
This will allow discrimination between
the various models suggested to explain each sub-class. The estimated
rate (per year) at which a given SN model could be distinguished from
alternative models is shown in the lower panel of Table 2.
The highest rate for discrimination between SN sub-classes 
with the INTERMEDIATE instrument (in either mode of 
operation) is less than one-SN per six years, thus those 
values are not shown.  These discrimination rates assume that
the explosion date is known to within $\pm$1 days from optical spectra
obtained as part of a coordinated study. 
It is unlikely that a SN will occur near enough for the 
the INTERMEDIATE instrument to be able to discriminate between 
explosion scenarios. The BASELINE instrument would be more successful, but 
would require a multi-year mission to accumulate more than a couple of 
distinguishable SNe Ia. The SUPERIOR ACT would be capable of distinguishing between 
the SN models we simulated for about 10 SNe Ia per year. Over a 5 year mission, for 
perhaps 30 normal SNe Ia, the SDCM and SC explosion scenarios could be distinguished. 
In addition, multiple SNe from the super- and sub-luminous sub-classes would be 
distinguishable, especially if the distance to the host galaxy were known. It is 
clear from these estimates that improved sensitivity and a large FoV are more useful 
for studying {\em prompt} SN emission than are improved angular and energy resolution.

\scriptsize
\begin{table}
\begin{center}
\caption{{\bf Detectability of Gamma-Ray Lines 
(5$\sigma$)
}}
\vspace{2mm}
\begin{tabular}{|l|ccccc|}
\hline
\footnotesize{SNR} & 
\multicolumn{5}{c|}{\footnotesize{Detectable Lines$^{a b c d}$}} \\
& {\bf SPI} & {\bf IBIS} & {\bf INTER} &
{\bf BASELINE} & {\bf SUPERIOR} \\
\hline
Cas A &T & &T & T & T,e+ \\
Tycho & 		  & & T,e+ & T,e+ & T,e+ \\
SN 1006 & e+ & & e+ & e+ & e+ \\
Vela &  & & e+ & a,e+,(F) & A,F,e+ \\
VelaJr.$^{e}$ & T,A,e+ &  & T,A,e+ & T,A,F,e+ & T,A,F,S,e+ \\
Kepler &  & & T & T & T,e+ \\
Cyg Lp & 	& & e+ & (A,e+,f) & A,F,e+ \\
Monoc &		&	 & e+ & (A,e+,f) &  A,F,e+ \\
LupusLp & 			 e+ & (e+) & e+ & e+,(F) & F,e+ \\
G6.5-12 &  e+ & (e+) & e+ & e+,(F) & F,e+ \\
CTB 13 &   & & e+ & (A,e+) & A,F,e+ \\
HB 21 & & & e+ -5yr & (A,e+)  &  A,e+,f \\
IC443 &				 & & & (e+ -5yr) & A,e+ \\
Crab &			 & & &  & T,A,e+ \\
Pup A & & & & (e+) & A,e+ \\
W44\protect$^{f}$ & & & & & e+ \\
SN1987A && & &  &  T,C \\
\hline
\end{tabular}
\small
\begin{tabular}{l}
\\
$^{a}$
T=$^{44}$Ti (1157 keV), 
A=$^{26}Al$ (1809 keV), e+ =511 keV, F=$^{60}$Fe (1173 keV), \\
C=$^{60}$Co (1173 keV), N=$^{22}$Na (1275 keV), S=$^{126}$Sn (666,695 keV) \\
$^{b}$The
$^{44}$Ti,$^{22}$Na \& $^{60}$Co lines were assumed to have 30 keV FWHM widths,
the 511, \\ 
Al26, Sn126, \& Fe60 
 lines were assumed to have 5 keV FWHM widths. \\
$^{c}$Lines in parentheses assume that all the emission emanates from 
a single compact knot. \\ 
Lines in lower-case require a 5 year mission.\\
$^{d}$ $^{22}$Na was not detected in any SN. \\
$^{e}$The remnant parameters for RXJ0852-4622 (Vela Jr). are currently debated. \\
$^{f}$The SNRs: RCW 103 \&  RCW 86 are similarly detectable. \\
\end{tabular}
\end{center}
\end{table}
\footnotesize

Using estimates of the distance, age, size and SN type of 19 SNRs
previously compiled in Milne (2000b) combined with 
estimates of the isotopic yields 
and the instrument specifications, we predict which gamma-ray lines will 
be detected from these SNRs (Table 3). Lines denoted by capital 
letters would be detected during a 2 year mission, while lines denoted by lower-case 
letter would require a 5 year mission. Lines encased in parantheses would only be 
detected if the single-knot approximation is valid. The INTERMEDIATE ACT would 
primarily be capable of studying positron annihilation radiation from nearby 
and/or type Ia remnants and of studying 1157 keV emission from young SNRs. The 
BASELINE and SUPERIOR ACTs would be capable of detecting line emission from a large 
number of remnants. Note that the Fe60 line would be detectable from up to six remnants 
with the BASELINE ACT, and from seven remnants with the SUPERIOR ACT.
 The BASELINE ACT performance would be greatly improved for 
clumpy emission. The SUPERIOR ACT would even detect emission from SN 1987A.

\vspace{-8mm}
\section{Discussion}

\vspace{-6mm}
It should suprise no one that improved specifications from an 
INTERMEDIATE to a BASELINE to a SUPERIOR ACT lead to 
predictions of increased  scientific capability. This exercise was 
performed not to demonstrate that the hypothetical SUPERIOR ACT 
is indeed a better ACT, but to quantify what can be anticipated 
from three different sets of specifications. The INTERMEDIATE 
ACT will not be an instrument ideally suited for studies of 
{\em prompt} emission from SNe Ia, but it will advance the 
understanding of galactic SNRs. It is up to the gamma-ray 
community to assess whether this level of gamma-ray SN science 
is important enough to include it as a selling point for an 
intermediate mission. 

Between the BASELINE and the SUPERIOR ACT designs, we have shown that
sensitivity and wide FoVs are crucial for high rates of SN detections  
and discrimination. The SUPERIOR ACT would be a truly breakthrough 
instrument for SN science, especially when consideration is given 
for the option of simultaneously 
using the SUPERIOR ACT in a second mode that accepts 
total deposition events (and thus improves the energy resolution and 
angular resolution for brighter objects). We argue that whether or not 
an intermediate ACT is constructed, the longer-term ACT is what is 
necessary for rich SN science. This ACT would be capable of 
studying SN rates, the explosion mechanism of type Ia SNe and the 
nucleosynthesis and ejecta kinematics in supernova remnants. 

\pagebreak

\end{document}